\documentstyle[prl,aps,epsfig,multicol,amssymb,amsfonts]{revtex}
\tighten
\renewcommand{\narrowtext} 
{\begin{multicols}{2}\global\columnwidth20.5pc} 
\renewcommand{\widetext}
{\end{multicols}\global\columnwidth42.5pc} 
\input psfig

\begin{document} 
\draft 
\title{Strong magnetoresistance induced by long-range disorder} 
\author{A.~D.~Mirlin,$^{1,*}$ J.~Wilke,$^{1}$ F.~Evers,$^1$
D.~G.~Polyakov,$^{2,\dagger}$ and P.~W\"olfle$^{1}$} 
\address{$^1$Institut f\"ur Theorie der Kondensierten Materie,
Universit\"at Karlsruhe, 76128 Karlsruhe, Germany}
\address{$^2$Institut
f\"ur Nanotechnologie, Forschungszentrum Karlsruhe, 76021 Karlsruhe,
Germany}
\date{\today}
\maketitle
\begin{abstract}
We calculate the semiclassical magnetoresistivity $\rho_{xx}(B)$ of
non-interacting
fermions in two dimensions
moving in a weak and smoothly varying random potential or random
magnetic field. We demonstrate that in a broad range of magnetic
fields the non-Markovian character of the transport leads to a strong
positive magnetoresistance. The effect is especially pronounced in the
case of a random magnetic field where $\rho_{xx}(B)$ becomes
parametrically much larger than its $B=0$ value. 
\end{abstract}

\narrowtext

The magnetoresistance (MR) is one of the most frequently studied
characteristics of the two-dimensional electron gas (2DEG).  
When the effect of disorder is
described by a collision integral within
the semiclassical Boltzmann equation approach,  
the resistivity tensor  $\hat\rho(B)$
for an isotropic system has the Drude form
\begin{equation}\label{rhodrude}
\hat\rho(B)=\frac{m}{e^2n}
\left(\begin{array}{cc} \tau^{-1} & \omega_c \\
-\omega_c & \tau^{-1}  \end{array}\right)\ ,
\end{equation}
where $n$ is the carrier density, $m$ the effective mass,
$\omega_c=eB/mc$ the cyclotron frequency, and $\tau$ the transport
scattering time. In particular, the longitudinal resistivity
$\rho_{xx}$ is independent of the magnetic field $B$,
$\rho_{xx}(B)=\rho_{0}\equiv m/e^2n\tau$, 
irrespective of the form of the impurity collision integral.
This result is solely
determined by the Markovian character of the transport assumed in the
Boltzmann equation description.

Deviations from the constant $\rho_{xx}(B)$ are conventionally termed
a positive/negative MR, depending on the sign of the deviation. The
negative MR \cite{lraa} induced by the suppression of the quantum
interference correction
by the magnetic field is a famous manifestation of 
weak localization.
Another source of negative MR is the Altshuler-Aronov correction
to the conductivity due to enhancement of the electron-electron
interaction by the diffusive motion of particles \cite{lraa}. Both
these effects are of quantum nature and lead to a correction of
order $e^2/h$ to the conductivity $\sigma_{xx}$, and thus to a
small correction to $\rho_{xx}$.

However, as we will show,
already at the classical level there exists a non-trivial MR
which can be much stronger than the quantum one, if the correlation
length $d$ of disorder is sufficiently large, $k_{F}d\gg 1$ (where
$k_{F}$ is the Fermi wave vector). This is due to memory effects
which  are neglected in the
collision integral description of disorder.

Transport properties of the 2DEG in a smooth random potential (RP)
$V({\bf r})$ are
of particular interest, since in currently fabricated high-mobility 
semiconductor heterostructures the disorder has long-range
character. The high mobility of these samples is achieved by placing
the charged donor ions in a layer separated by a large distance $d$
($k_F d\sim 10$) from the 2DEG plane. Assuming the positions of these
impurities to be statistically distributed with a sheet density $n_i$,
the correlation function $W_{V}({\bf r}-{\bf r}')=\langle
V({\bf r})V({\bf r}') \rangle$ is given in 
momentum space by 
\begin{equation}\label{wv}
\tilde W_{V}(q)=(\pi\hbar^2/m)^2 n_i e^{-2qd}\ .
\end{equation}

A new type of transport problem occurs in these systems when a large
magnetic field $B\simeq B_{1/2}=2(hc/e)n$ is applied such that the
lowest Landau level is approximately half filled. The metallic state
then observed has been described \cite{halperin93} in
terms of composite fermions (CF's) moving in a weak effective field $\bar
B=B-B_{1/2}$. The CF's are scattered by an impurity-induced random
magnetic field (RMF) $\delta B({\bf r})$ characterized by
 the correlation function $W_{B}({\bf r}-{\bf r}')=\langle
\delta B({\bf r})\delta B({\bf r}') \rangle$ with Fourier components
\begin{equation}\label{wb}
\tilde W_{B}(q)=(2h c /e)^2 n_i e^{-2qd}\ .
\end{equation}
While the above RMF is fictitious,
a real long-range correlated RMF can also be realized in
semiconductor heterostructures by attaching superconducting
\cite{bending90,geim92} or ferromagnetic \cite{ye96,zielinski98}
overlayers. We will study below the case of a weak RP or RMF, which
means that $l\gg d$, where $l$ is the mean free path in zero average
field $\bar B=0$.  Let us stress that we consider a situation with
only the smooth disorder (RP or RMF) present. This should be
contrasted with the starting point of \cite{hs}, where the resistivity
was assumed to be dominated by a white-noise RP while a weak
long-range RMF was considered as a small perturbation.

As one manifestation of the strongly non-Markovian transport, it has
been shown 
recently \cite{fogler97,mirlin98,empw} that in a sufficiently strong
$\bar B$ 
the MR drops exponentially with $\bar B$ because of a ``classical
localization'' effect caused by adiabacity of the motion. This holds
true both for the motion in a RP \cite{fogler97} and in a RMF
\cite{mirlin98,empw}. The condition of this adiabatic regime is
$\omega_c\tau\gg (l/d)^{2/3}$.

In this paper we study the region of smaller magnetic fields in which
different non-Markovian processes become important. We
find that the exponential falloff of $\rho_{xx}$ is preceded
by a strong positive MR. The effect is especially pronounced in the
case of the RMF, where the increase of $\rho_{xx}$ is much larger (in
the weak disorder limit) than its zero-$\bar B$ value $\rho_{0}=
m/e^2n\tau$.

We first outline the physics of the effect on a qualitative
level. The zero-$\bar B$ transport scattering rates in the RP and RMF
are given by \cite{maw}
\begin{eqnarray}
\frac{1}{\tau}
 &=&\frac{1}{2\pi
m^2v_{F}^3}\int_0^{\infty}\!\!dq\;q^2 \tilde{W}_{V}(q) \quad \mbox{(RP)};
\label{tauv} \\ 
\frac{1}{\tau}
 &=&\left(\frac{e}{mc}\right)^2
\frac{1}{2\pi v_{F}}\int_0^{\infty}\!\!dq\;\tilde{W}_{B}(q) 
\quad \mbox{(RMF)}. \label{taub}
\end{eqnarray}
Let us now discuss the nature of the particle trajectories. The change
$d\phi$ of the polar angle of the particle velocity in the time
interval $dt$ is given by
\begin{eqnarray} \label{dphiv}
d\phi&\simeq& (mv_{F})^{-1} n_i \epsilon_{ij} \partial_{j} V({\bf r})
dt \quad \mbox{(RP);} \\
\label{dphib}
d\phi&=& (e/mc)\delta B({\bf r}) dt \quad \mbox{(RMF),}
\end{eqnarray}
where ${\bf n} = {\bf v}/|{\bf v}|$ is the unit vector in the direction
of the velocity. Taking into account that the trajectory within the
correlation domain of the size $\sim d$ is almost a straight line, one
gets (\ref{tauv}), (\ref{taub}). One source of the MR is the bending
of the trajectory within the correlation domain by the magnetic field
$\bar B$. This leads to a small negative MR, of the relative magnitude
$\sim (d/R_c)^2\ll 1$ \cite{khveshchenko96,mirlin98}. There exists,
however, a much stronger effect related to returns of the particle to
spatial regions close to the starting point. Under the condition
$\omega_c\tau\gg 1$ the particle trajectory is a sequence of slightly
distorted cyclotron circles. The center of the orbit is shifted by a
random vector $\bbox{\delta}$ after one cyclotron revolution, with
$\langle\bbox{\delta}\rangle=0$,
$\langle\delta_{x}^2\rangle=\langle\delta_{y}^2\rangle=2\pi
l^2/(\omega_c\tau)^3$. 
If, for definiteness, the velocity ${\bf v}\|{\bf \hat x}$ at the time
$t=0$, the trajectory passes a distance $\delta_y$ from the original
point at the time $t\simeq 2\pi/\omega_c$. The correlation of 
the scattering processes (\ref{dphiv}), (\ref{dphib})
at $t\simeq 0$ and
$t\simeq 2\pi/\omega_c$ induced by the long-range character of the
disorder leads to a correction to the relaxation rate and, therefore,
the resistivity 
\begin{eqnarray} \label{d1v}
(\Delta\rho_{xx}/\rho_0)_1 &\sim [d/\langle \bbox{\delta}^2\rangle ^{1/2}]^3
\quad \mbox{(RP);} \\ \label{d1b}
(\Delta\rho_{xx}/\rho_0)_1 &\sim d/\langle \bbox{\delta}^2\rangle ^{1/2} \quad
\mbox{(RMF).} 
\end{eqnarray}

The same is valid for $t\simeq2\pi n/\omega_c$ (the $n$-th cyclotron
revolution, $n=2,3,\dots$), with $\langle \bbox{\delta}^2\rangle$
multiplied by $n$. For the RP case the corresponding sum over $n$
converges, $\sum_{n=1}^{\infty}n^{-3/2}=\zeta(3/2)$, leading simply to the
renormalization of a numerical factor in (\ref{d1v}). In fact, in this
case the above consideration can be made fully quantitative, and the
result agrees with what we will find below from the
Liouville equation (\ref{lio}). The corresponding positive MR, $\Delta
\rho_{xx}/\rho_0\sim (d/l)^3(\omega_c\tau)^{9/2}$, becomes of order unity
at the upper bound of the considered range of the magnetic fields,
$\omega_c\tau \lesssim(l/d)^{2/3}$. Here the system enters the
adiabatic regime.

In the RMF case the sum over $n$ is of the form $\sum n^{-1/2}$ and is
thus determined by the upper cutoff, which is $n\sim
\omega_c\tau$. The resulting correction
$\Delta\rho_{xx}/\rho_0\sim(d/l)(\omega_c\tau)^2$ reaches a value of order 
unity at $\omega_c\tau\sim(l/d)^{1/2}$, i.e. {\it far from the adiabatic
regime}. At larger magnetic fields, where $\Delta\rho_{xx}/\rho_{0}\gg
1$, a self-consistent treatment is needed (see below).

We describe now a formalism which allows us to calculate the
MR more systematically. We consider first the RMF case. The
starting point is the Liouville equation
\begin{eqnarray}\label{lio}
(L_0+\delta L)g(\omega,{\bf r},\phi)&=&\cos{(\phi-\phi_E)}; \\ \nonumber
L_0=-i\omega+v_{F}{\bf n}\nabla+\frac{e}{mc}\bar B
\frac{\partial}{\partial \phi}&;& \quad \delta L = \frac{e}{mc}\delta B ({\bf
r})\frac{\partial}{\partial \phi} 
\end{eqnarray}
for the deviation $\delta f(\omega,{\bf r},\phi)=eEv_{F}\frac{\partial
f_0}{\partial \epsilon} g(\omega,{\bf r},\phi)$ from the equilibrium
distribution function $f_0=\theta(\epsilon_F-\epsilon)$. Here ${\bf
E}=E(\cos{\phi_E}, \sin{\phi_E})$ is the electric field and ${\bf
n}=(\cos{\phi}, \sin{\phi})$ the unit vector determining the velocity
direction. The current density is given by ${\bf
j}=-e\int\frac{d^2p}{(2\pi\hbar)^2} {\bf v} \delta f$, which yields
the conductivity tensor
\begin{equation}\label{stensor}
\hat \sigma = e^2v_{F}^2N_F \!\int \! \frac{d\phi}{2\pi} \left\langle\!
\left(\begin{array}{c} \cos{\phi}\\ 
                        \sin{\phi}\end{array}\right)
(L_0+\delta L)^{-1} \!
\left(\begin{array}{c} \cos{\phi}\\ 
                       \sin{\phi}\end{array}\right)^T\right\rangle ,
\end{equation}
where $N_F$ is the density of states and the angular brackets denote
the averaging over configurations of the RMF $\delta B({\bf r})$ with
the correlation function (\ref{wb}).

Expanding (\ref{stensor}) in $\delta L$, averaging over the disorder
and resumming the series (in the same way as it is done for a
quantum-mechanical Green's function), one arrives at
\begin{eqnarray}\label{rtensor}
\hat \rho &=& \hat \sigma ^{-1} = \frac{2}{e^2 v_{F}^2 N}(\hat L_0 +
\hat M); \\
&&\hat L_0=\left(\begin{array}{cc} -i\omega & \omega_c \\ \nonumber
-\omega_c & -i\omega \end{array}\right),
\end{eqnarray}
where $M$ is the ``self-energy''
and the $2\times 2$ matrix  $\hat M$ is defined by
\begin{equation}\label{opdef}
\hat M=\left\langle\begin{array}{c} \cos{\phi}\\\sin{\phi}\end{array}
\right| M \left| 
\begin{array}{c} \cos{\phi}\\\sin{\phi}\end{array}\right\rangle\ .
\end{equation}

In zero $\bar B$ the self-energy $M$ is given in the leading
approximation by the first term of the perturbative expansion,
$M=-\langle\delta L \; L_0^{-1}\; \delta L \rangle$, yielding
\begin{eqnarray}\nonumber
M_{xx}&=&-2i\left(\frac{e}{mc}\right)^2
\int\!\frac{d^2q}{(2\pi)^2}\frac{d\phi}{2\pi}\;
 \\ \label{mtau}
&\times & \sin{\phi}\frac{\tilde{W}_{B}(q)}
{v_{F}q\cos{(\phi-\phi_{q})}-\omega} \sin{\phi}\ ,
\end{eqnarray}
where $\phi_q$ is the polar angle of the momentum ${\bf q}$. Taking
into account that $\omega$ should have an infinitesimal positive
imaginary part ($\omega\rightarrow \omega + i0$) and considering the
limit $\omega \rightarrow 0$, we get $M_{xx}=1/\tau$ with $1/\tau$
given by (\ref{taub}). Equation (\ref{rtensor}) reproduces then the
Drude resistivity, as expected.

Now we calculate the $\bar B$-dependent correction to (\ref{mtau})
determined by the return processes described above. For this purpose,
we have to replace the free propagator
$L_0^{-1}=(-i\omega+iv_{F}q\cos{(\phi-\phi_q)})^{-1}$ entering
(\ref{mtau}) by the one describing the motion in the magnetic field in
the presence of disorder. At small $q$ this would be simply the
diffusion propagator; however, since we are interested in the
short-scale physics, $q \gtrsim R_c^{-1}$, the diffusion approximation
is not appropiate. Using the fact that the particle is
only scattered by a small angle within the correlation length $d$, we
can approximate the motion by a Fokker-Planck equation corresponding
to the diffusion {\it in momentum space}:
\begin{eqnarray}\nonumber
&&\Delta M_{xx}=2\left(\frac{e}{mc}\right)^2\int\!\frac{d^2q}{(2\pi)^2}\;
\tilde{W}_{B}(q) \\ \label{mbb}
&&\qquad\times\int\!\frac{d\phi}{2\pi}\; 
\sin{\phi}\: g_{\rm D}(\omega, {\bf q}, \phi),
\end{eqnarray}
where $g_{\rm D}$ is the solution of 
\begin{eqnarray}
 \nonumber
&&\left[-i\omega+ivq\cos{(\phi-\phi_q)}+\omega_c\frac{\partial}{\partial
\phi}-\frac{1}{\tau}\frac{\partial^2}{\partial \phi^2}\right] 
g_{\rm D}(\omega, {\bf q}, \phi)\\
\label{fokplanck}
&& \qquad =\sin{\phi}\ .
\end{eqnarray}
Now we make use of the fact that the solution of
(\ref{fokplanck}) determines the spatial dispersion
of the conductivity in the situation of  small-angle scattering
\cite{mw-prl,mw-prb}: 
\begin{equation}\label{syy0}
\sigma_{yy}(\omega, {\bf q})=e^2 v_F^2 N_F
\int\!\frac{d\phi}{2\pi}\;\sin{\phi}\:g_{\rm D}(\omega, {\bf q}, \phi)\ .
\end{equation}
At $\omega=0$ one has $\sigma_{yy}({\bf q}\| {\bf\hat y})=0$
\cite{mw-prl}. For ${\bf q}\| {\bf\hat x}$ we get, using the solution
of (\ref{fokplanck}) at $\omega_c\tau \gg 1$ \cite{mw-prb},
\begin{eqnarray}\nonumber
\sigma_{yy}(0, {\bf
q})&\simeq&\frac{2\sigma_0}{\omega_c\tau}\int_0^{2\pi}\!
\frac{d\phi}{2\pi}\int_{-\infty}^{\phi}\!d\phi'\;
\sin{\phi}\sin{\phi'}e^{-K(\phi,\phi')}
\\ \label{syy1}
&\simeq&4\sigma_0J_1^2(qR_c)/(qR_c)^2\ ,
\end{eqnarray}
where
$K(\phi,\phi')=iqR_c(\sin{\phi}-\sin{\phi'})-
[(qR_c)^2/2\omega_c\tau](\phi-\phi')$ and $\sigma_0=\rho_0^{-1}$.   
This determines the integrand of (\ref{mbb}) for general orientation
of ${\bf q}$, with the result for $\omega=0$
\begin{eqnarray}\nonumber
\Delta M_{xx}&=&\left(\frac{e}{mc}\right)^2\frac{\tau}{\pi
R_c^2}\int_0^{\infty}\!\!\frac{dq}{q}\; J_1^2(qR_c)\tilde W_B(q)\\ \label{mbb1}
&\simeq&\left(\frac{e}{mc}\right)^2\frac{\tau}{2\pi R_c^2}\tilde
W_B(0)\ .
\end{eqnarray}
In agreement with the above qualitative picture, the main contribution
to the integral comes from $q\sim R_c^{-1}$. 
Therefore, we neglected the {\it q}-dependence of
$\tilde W_B(q)$ in the second line of (\ref{mbb1}), in view of $d\ll
R_c$. The MR is thus positive and quadratic in $\bar B$,
\begin{equation}\label{rhoxxbb}
\Delta\rho_{xx}/\rho_0=(\bar B/B_0)^2\equiv 
4\alpha^2(\omega_c\tau)^2\equiv2(d/l)(\omega_c\tau)^2,
\end{equation}
where $B_0=\langle\delta B^2({\bf r})\rangle^{1/2}\equiv W_B^{1/2}(0)$
is the amplitude of the RMF fluctuations and $\alpha\ll 1$ is the
parameter characterizing the RMF strength \cite{mirlin98,empw},
$\alpha=d\omega_c^{(0)}/v_{F}$ with
$\omega_c^{(0)}=eB_0/mc$. Eq.~(\ref{rhoxxbb}) 
is valid for $\bar B\ll B_0$, whereas the adiabatic regime begins at
$\bar B\sim B_0\alpha^{-1/3}$. In the intermediate range,
$B_0\lesssim\bar B\lesssim B_0\alpha^{-1/3}$, the positive MR gets
large, $\Delta\rho_{xx}/\rho_0\gg 1$. In this region, Eq.~(\ref{mbb1})
should be treated self-consistently, i.e. $\tau$ in the r.h.s. should
be understood as a renormalized scattering time,
$\tau^{-1}=\tau_0^{-1}+\Delta M_{xx}$. The result for $\bar B \gg B_0$ is
$\rho_{xx}/\rho_0= \bar B/B_0$, or, in the form valid both below
and above $B_0$,
\begin{equation}\label{rhoxxsc}
\rho_{xx}/\rho_0=1/2+[1/4+(\bar B/B_0)^2]^{1/2}\ ,
\end{equation}
which is our main result for the case of RMF.
At $\bar B/B_0\sim \alpha^{-1/3}$ the resistivity reaches its maximum
$\rho_{xx}\sim\alpha^{-1/3}\rho_0$; in still higher fields 
$\rho_{xx}$ drops rapidly due to the adiabatic character of motion.

We turn now to the RP scattering. The operator $\delta L$ in
(\ref{lio}) has then the form
\begin{equation}\label{deltalv}
\delta L=\delta v({\bf r}){\bf n}\nabla+(\nabla\delta v({\bf r})){\bf
n}_{\perp}\frac{\partial}{\partial \phi},
\end{equation}
where $\delta v({\bf r})=V({\bf r})/p_{F}$ is the spatial variation of
the Fermi velocity and ${\bf n}_{\perp}={\bf \hat z}\times {\bf
n}=(-\sin{\phi}, \cos{\phi})$. At zero $\bar B$ we get, instead of
(\ref{mtau}),
\begin{eqnarray}\nonumber
M_{xx}&=&-\frac{2i}{p_{F}^2}\int\!
\frac{d^2q}{(2\pi)^2}\frac{d\phi}{2\pi}\;\sin{\phi}\sin{(\phi-\phi_q)}
\\ \label{mtauv}
&\times& \frac{q^2\tilde{W}_{V}(q)}{v_{F}q
\cos{(\phi-\phi_{q})}-\omega}\sin{\phi}\sin{(\phi-\phi_q)},
\end{eqnarray}
reproducing the result (\ref{tauv}). To calculate the MR, we replace,
as in the RMF case, the free propagator in (\ref{mtauv}) by
$L_{\rm D}^{-1}$, where $L_{\rm D}$ is the operator in the square brackets in
(\ref{fokplanck}). Solving again the equation $L_{\rm D}g_{\rm D}(\omega, {\bf
q}, \phi)=\sin{\phi}\sin{(\phi-\phi_q)}$ at $\omega_c\tau\gg1$,
we find after some algebraic manipulations
$$
\Delta M_{xx}\simeq \frac{1}{2\pi p_{F}^2
v_{F}}\int_0^{\infty}\!\!dq\;q^2\tilde W_{V}(q)
\left[\coth{\frac{\pi(qR_c)^2}{2\omega_c\tau}}-1\right].
$$
In contrast to (\ref{mbb1}),
the integral is determined by the region of momenta $q^2\sim
\omega_c\tau/R_c^2\sim \langle \bbox{\delta}^2\rangle^{-1}$, as might be
expected from the above qualitative consideration. Since $d\ll\langle
\bbox{\delta}^2\rangle^{1/2}$, we can again neglect the $q$-dependence of
$\tilde W_{V}(q)$, which yields for $\omega_c\tau\ll (l/d)^{2/3}$
\begin{eqnarray}\nonumber
\frac{\Delta\rho_{xx}}{\rho_0}&=&\frac{\zeta(3/2)}{4\pi^2}\frac{\tau
\tilde W_{V}(0)}{p_{F}^2v_{F}l^3}(\omega_c\tau)^{9/2}\\\label{rhoxxbv}
&=&\frac{2\zeta(3/2)}{\pi}\left(\frac{d}{l}\right)^3(\omega_c\tau)^{9/2}.
\end{eqnarray}

\begin{figure}
\includegraphics[width=0.95\columnwidth,clip]{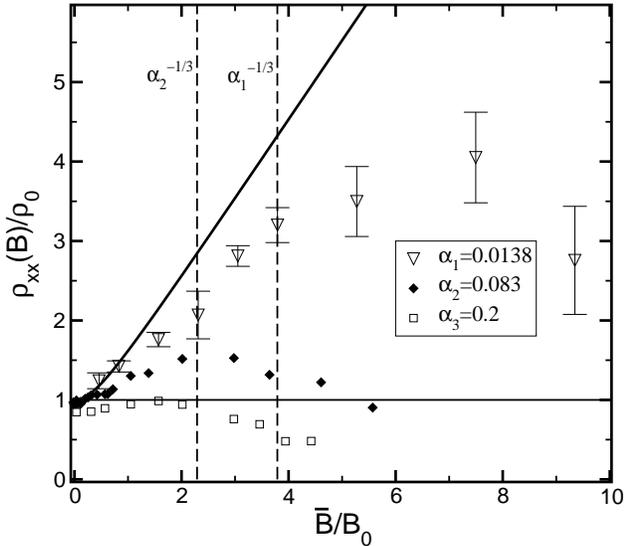}
\caption{Magnetoresistivity [normalized by the Drude value
$\rho_0=m/e^2n\tau$ with $1/\tau$ given by (\ref{taub})] 
in random magnetic field from the numerical
simulations for three
different strengths of the disorder; the full line corresponds to 
Eq.~(\ref{rhoxxsc}).}
\label{rmfres} 
\end{figure}

We have performed numerical simulations of the MR for both types
of disorder (RMF and RP). In Fig.~\ref{rmfres} the results for the RMF
are shown, for three different strengths of the disorder ($\alpha=0.2$,
0.083, 0.0138). At $\alpha\ll 1$, the theoretical prediction of the
strong positive MR (\ref{rhoxxsc}) crossing over to the negative one at
$\bar B\sim B_0\alpha^{-1/3}$ is fully confirmed by the data. At
moderately small $\alpha$ ($\alpha=0.2$ in Fig.~\ref{rmfres}) the positive
MR still exists, but becomes weak; this is the region of $\alpha$
relevant to the composite-fermion description of the vicinity of half
filling of the lowest Landau level ($\nu=1/2$). The numerically
calculated MR for $\alpha\sim 0.2-0.3$ agrees well \cite{empw} with
the experimental data around $\nu=1/2$. At large $\alpha\geq 0.5$ the
region of positive MR disappears, and $\rho_{xx}$ drops
monotonously with $\bar B$ \cite{empw}.

\vspace{-3mm}

\begin{figure}
\includegraphics[width=0.95\columnwidth,clip]{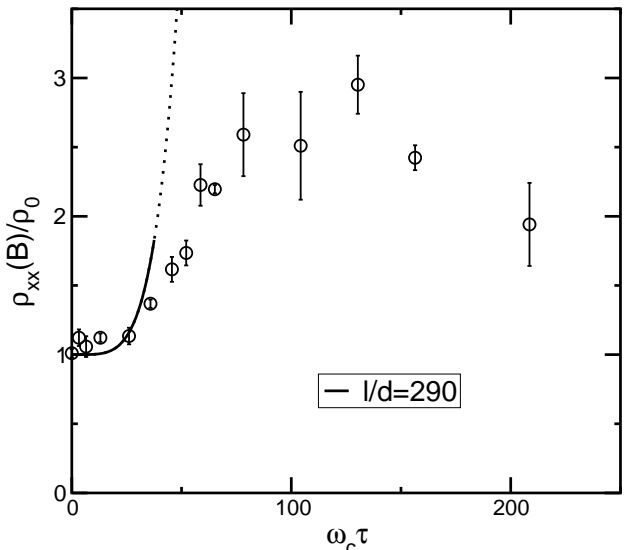}
\caption{Magnetoresistivity in a random potential from computer simulations in
comparison with Eq.~(\ref{rhoxxbv}).  }
\label{rpres} 
\end{figure}

The numerically found MR for the RP case (Fig.~\ref{rpres}) shows good
agreement with the theoretical result (\ref{rhoxxbv}) up to $\Delta
\rho_{xx}/\rho_0 \sim 1$. At larger $\bar B$, $\rho_{xx}$
deviates from (\ref{rhoxxbv}) and starts to decrease, as expected.

In conclusion, we have demonstrated that the 2D fermion gas shows for
 weak long-range correlated disorder a strong positive MR in
moderately strong magnetic fields $1\ll\omega_c\tau\lesssim (l/d)^{2/3}$,
due to the non-Markovian character of transport. The effect is
especially pronounced in the case of the RMF. Our findings
explain, in particular, the positive MR of composite fermions observed
experimentally around $\nu=1/2$.

This work was supported by the SFB 195 der Deutschen
Forschungsgemeinschaft and by the INTAS grant No. 97-1342.

\end{multicols}
\end{document}